\documentclass[aps,prb,twocolumn,showpacs]{revtex4}

\usepackage{graphicx}
\usepackage{amsmath}
\usepackage{amssymb}

\begin{document}

\title{Pseudogap and its connection to particle-hole asymmetry electronic state and Fermi arcs in cuprate
superconductors}


\author{Huaisong Zhao, L\"ulin Kuang, and Shiping Feng$^{*}$}
\affiliation{Department of Physics, Beijing Normal University,
Beijing 100875, China}

\begin{abstract}
The particle-hole asymmetry electronic state of cuprate superconductors and the related doping and temperature
dependence of the Fermi arc length are studied based on the kinetic energy driven superconducting mechanism. By taking
into account the interplay between the SC gap and normal-state pseudogap, the essential feature of the evolution of the
Fermi arc length with doping and temperature is qualitatively reproduced. It is shown that the particle-hole asymmetry
electronic state is a natural consequence due to the presence the normal-state pseudogap in the particle-hole channel.
The Fermi arc length increases with increasing temperatures below the normal-state pseudogap crossover temperature
$T^{*}$, and it covers the full length of the Fermi surface for $T>T^{*}$. In particular, in analogy to the temperature
dependence of the Fermi arc length, the low-temperature Fermi arc length in the underdoped regime increases with
increasing doping, and then it evolves into a continuous contour in momentum space near the end of the superconducting
dome. The theory also predicts an almost linear doping dependence of the Fermi arc length.
\end{abstract}


\maketitle

\section{Introduction}

In the conventional superconductors \cite{schrieffer83}, an energy gap exists in the electronic energy spectrum only
below the superconducing (SC) transition temperature $T_{\rm c}$, which is corresponding to the energy for breaking a
Cooper pair of the charge carriers and creating two excited states. However, in cuprate superconductors above
$T_{\rm c}$ but below a temperature $T^{*}$, an energy gap called the normal-state pseudogap exists
\cite{Ding96,Batlogg94,Hufner08,Timusk99}. Although the SC gap has a domelike shape of the doping dependence
\cite{Damascelli03}, the magnitude of the normal-state pseudogap is much larger than that of the SC gap in the
underdoped regime \cite{Ding96,Batlogg94,Hufner08,Timusk99}, then it smoothly decreases upon increasing doping, and
seems to merge with the SC gap in the overdoped regime, eventually disappearing together with superconductivity at the
end of the SC dome \cite{Hufner08}. Since many of the unusual physical properties of cuprate superconductors have often
been attributed to particular characteristics of the low-energy excitations determined by the electronic structure
\cite{Hufner08,Timusk99,Damascelli03}, the normal-state pseudogap observed in the excitation spectrum as a suppression
of the spectral weight is thought to be key to understanding the mechanism of superconductivity.

During the last two decades, the angle-resolved photoemission spectroscopy (ARPES) has been emerged as a powerful tool
for studying the electronic structure of cuprate superconductors, since it is a direct method for probing the momentum
dependence of the SC gap and the locus in the momentum space where the quasiparticle excitations are gapless
\cite{Damascelli03}. In spite of the nonconventional SC mechanism, the ARPES experimental results have unambiguously
established the Bogoliubov-quasiparticle nature of the sharp SC quasiparticle peak in cuprate superconductors in the
overdoped regime \cite{Lee07}, then the SC coherence of the low energy quasiparticle excitation is well fitted by a
simple Bardeen-Cooper-Schrieffer (BCS) formalism \cite{schrieffer83} with a d-wave SC gap. However, the pseudogap state
is particularly obvious in the underdoped regime \cite{Ding96,Batlogg94,Hufner08,Timusk99}, this leads to that the
physical properties of cuprate superconductors in the underdoped regime exhibit a number of the anomalous properties \cite{Batlogg94,Hufner08,Timusk99,Damascelli03,Norman98,Kanigel06,Nakayama09,Kondo09,Meng11}. In particular, the ARPES
experimental data show that in the underdoped regime, although the normal-state of cuprate superconductors is metallic,
the part of the Fermi surface is gapped out by the normal-state pseudogap, then the low-energy electron excitations
occupy disconnected segments called as the Fermi arcs located at the nodal region in the Brillouin zone
\cite{Norman98,Kanigel06}. Moreover, in corresponding to the doping and temperature dependence of the normal-state
pseudogap, the essential feature of the evolution of the Fermi arc length with doping and temperature has been
established now \cite{Norman98,Kanigel06,Nakayama09,Kondo09,Meng11}: (1) the Fermi arc in the underdoped regime
increases in length with temperatures, till at about the normal-state pseudogap crossover temperature $T^{*}$, then
it covers the full length of the Fermi surface (a continuous contour in momentum space) for the temperature $T>T^{*}$
\cite{Norman98,Kanigel06,Nakayama09}; (2) the Fermi arc increases its length as a function of doping \cite{Meng11},
and then it evolves into a continuous contour in momentum space near the end of the SC dome. In particular, the
experimental data indicate both the particle-hole symmetry breaking and the pronounced spectral broadending due to the
presence of the normal-state pseudogap, which reflect the spatial symmetry breaking without long-range order at the
opening of the normal-state pseudogap \cite{Hashimoto10,He11,Daou10}. It is thus established that the low-energy
quasiparticle excitations at the Fermi energy dramatically change with doping and temperature, and have a close
relation to the normal-state pseudogap.

Although the main features of the particle-hole asymmetry electronic state in cuprate superconductors and the related
doping and temperature dependence of the Fermi arc length are well-established by now
\cite{Norman98,Kanigel06,Nakayama09,Kondo09,Meng11,Hashimoto10,He11,Daou10}, their full understanding is still
a challenging issue. In our recent work \cite{feng12}, the interplay between the SC gap and normal-state pseudogap in
cuprate superconductors has been studied based on the kinetic energy driven SC mechanism \cite{feng0306}, where the
charge carriers interact directly through the kinetic energy by exchanging spin excitations, then this microscopic
interaction provides a natural explanation of both the origin of the normal-state pseudogap state in the particle-hole
channel and the pairing mechanism for superconductivity in the particle-particle channel \cite{feng12}. In this paper,
we study the low-energy electronic structure of cuprate superconductors along with this line \cite{feng12}. We evaluate
explicitly the electron spectral function by taking into account the interplay between the SC gap and normal-state
pseudogap, and then qualitatively reproduce the main features of the ARPES measurements on cuprate superconductors
\cite{Norman98,Kanigel06,Nakayama09,Kondo09,Meng11,Hashimoto10,He11}, including the doping and temperature dependence
of the Fermi arc length.

The rest of this paper is organized as follows. We present the basic formalism in Section \ref{framework}, while the
quantitative characteristics of the particle-hole asymmetry electronic state are discussed in Section
\ref{electronic-structure}, where we show that the particle-hole asymmetry electronic state and the related doping and
temperature dependence of the Fermi arc length are intriguingly related to the emergence of the normal-state pseudogap
in the particle-hole channel. Finally, we give a summary in Section \ref{conclusions}.


\section{Theoretical framework}\label{framework}

In cuprate superconductors, the single common feature is the presence of the CuO$_{2}$ plane \cite{Damascelli03},
and it seems evident that the unusual behavior is dominated by this plane. Very soon after the discovery of
superconductivity in cuprate superconductors, it has been argued that the essential physics of the doped CuO$_{2}$
plane is contained in the $t$-$J$ model on a square lattice \cite{anderson87},
\begin{eqnarray}\label{t-J-model}
H&=&-t\sum_{i\hat{\eta}\sigma}C^{\dagger}_{i\sigma}C_{i+\hat{\eta}\sigma}+t'\sum_{i\hat{\tau}\sigma}
C^{\dagger}_{i\sigma}C_{i+\hat{\tau}\sigma}\nonumber\\
&+&\mu\sum_{i\sigma} C^{\dagger}_{i\sigma}C_{i\sigma}+J\sum_{i\hat{\eta}}{\bf S}_{i}\cdot
{\bf S}_{i+\hat{\eta}},
\end{eqnarray}
supplemented by an important on-site local constraint $\sum_{\sigma}C^{\dagger}_{l\sigma}C_{l\sigma}\leq 1$ to remove
the double occupancy, where the summation is over all sites $i$, and for each $i$, over its nearest-neighbors
$\hat{\eta}$ or the next nearest-neighbors $\hat{\tau}$, $C^{\dagger}_{i\sigma}$ and $C_{i\sigma}$ are electron
operators that respectively create and annihilate electrons with spin $\sigma$,
${\bf S}_{i}=(S^{\rm x}_{i},S^{\rm y}_{i},S^{\rm z}_{i})$ are spin operators, and $\mu$ is the chemical potential. For
a proper treatment of the electron single occupancy local constraint in the $t$-$J$ model (\ref{t-J-model}), a
charge-spin separation (CSS) fermion-spin theory \cite{feng04,feng08} has been proposed, where a spin-up annihilation
(spin-down annihilation) operator for the physical electron is given by a composite operator as
$C_{i\uparrow}=h^{\dagger}_{i\uparrow}S^{-}_{i}$ ($C_{i\downarrow}=h^{\dagger}_{i\downarrow}S^{+}_{i}$), with the
spinful fermion operator $h_{i\sigma}=e^{-i\Phi_{i\sigma}}h_{i}$ that keeps track of the charge degree of freedom of
the electron together with some effects of spin configuration rearrangements due to the presence of the doped hole
itself (charge carrier), while the spin operator $S_{i}$ describes the spin degree of freedom of the electron, then
the electron single occupancy local constraint is satisfied in analytical calculations. In this CSS fermion-spin
representation, the $t$-$J$ model (\ref{t-J-model}) can be rewritten as \cite{feng04,feng08},
\begin{eqnarray}\label{css-t-Jmodel}
H&=&t\sum_{i\hat{\eta}}(h^{\dagger}_{i+\hat{\eta}\uparrow}h_{i\uparrow}S^{+}_{i}S^{-}_{i+\hat{\eta}}+
h^{\dagger}_{i+\hat{\eta}\downarrow}h_{i\downarrow}S^{-}_{i}S^{+}_{i+\hat{\eta}})\nonumber\\
&-&t'\sum_{i\hat{\tau}}
(h^{\dagger}_{i+\hat{\tau}\uparrow}h_{i\uparrow}S^{+}_{i}S^{-}_{i+\hat{\tau}}+h^{\dagger}_{i+\hat{\tau}\downarrow}
h_{i\downarrow}S^{-}_{i}S^{+}_{i+\hat{\tau}})\nonumber\\
&-&\mu\sum_{i\sigma} h^{\dagger}_{i\sigma}h_{i\sigma}+J_{{\rm eff}}\sum_{i\hat{\eta}}{\bf S}_{i}
\cdot {\bf S}_{i+\hat{\eta}},
\end{eqnarray}
where $J_{{\rm eff}}=(1-\delta)^{2}J$, and the doping concentration
$\delta=\langle h^{\dagger}_{i\sigma}h_{i\sigma}\rangle=\langle h^{\dagger}_{i}h_{i}\rangle$.

As in the conventional superconductors \cite{schrieffer83}, the SC-state in cuprate superconductors is also
characterized by the Cooper pairs, forming SC quasiparticles \cite{tsuei00}. However, as a natural consequence of the
unconventional SC mechanism that is responsible for the high SC transition temperatures \cite{anderson87}, the Cooper
pair in cuprate superconductors has a dominant d-wave symmetry \cite{tsuei00,Damascelli03}. In this case, one of the
main concerns in the field of superconductivity in cuprate superconductors is about the origin of the d-wave Cooper
pairs. From the experimental side, it has been well established that the antiferromagnetic short-range correlation
coexists with the SC-state in the whole SC regime \cite{yamada98,dai01,Stock05,arai99,Bourges05}, which provides a
clear link between the pairing mechanism and magnetic excitation. On the theoretical hand, we have developed a kinetic
energy driven SC mechanism \cite{feng0306} based on the $t$-$J$ model (\ref{css-t-Jmodel}), where the charge carrier
interaction directly from the kinetic energy by exchanging spin excitations induces a d-wave charge carrier pairing
state in the particle-particle channel, then the electron Cooper pairs originating from the charge carrier pairing
state are due to the charge-spin recombination, and their condensation reveals the SC ground-state. Moreover, this
SC-state is controlled by both the SC gap and quasiparticle coherence, which leads to that the maximal SC transition
temperature occurs around the optimal doping, and then decreases in both the underdoped and overdoped regimes. This
microscopic SC theory gives a consistent description of the physical properties of cuprate superconductors, including
the doping and temperature dependence of the microwave conductivity \cite{wang08}, the extinction of the quasiparticle
scattering interference \cite{wang10}, and the doping dependence of the Meissner effect \cite{feng10}. In particular,
it has been shown recently that besides the pairing mechanism in the particle-particle channel provided by the charge
carrier interaction directly from the kinetic energy by exchanging spin excitations, this same microscopic interaction
also induces the normal-state pseudogap state in the particle-hole channel \cite{feng12}. Based on this work
\cite{feng12}, we \cite{zhao12} have discussed the doping dependence of the specific-heat of cuprate superconductors,
and shown that the striking behavior of the specific-heat humplike anomaly near $T_{\rm c}$ in the underdoped regime
can be attributed to the emergence of the normal-state pseudogap. Following our previous discussions
\cite{feng0306,feng12}, the full charge carrier diagonal and off-diagonal Green's functions of the $t$-$J$ model
(\ref{css-t-Jmodel}) in the SC-state can be obtained as,
\begin{widetext}
\begin{subequations}\label{hole-Green's-function}
\begin{eqnarray}
g({\bf k},\omega)&=&{1\over \omega-\xi_{\bf k}-\Sigma^{({\rm h})}_{1}({\bf k},\omega)-\bar{\Delta}^{2}_{\rm h}({\bf k})
/[\omega+\xi_{\bf k}+\Sigma^{({\rm h})}_{1}({\bf k},-\omega)]},\label{hole-diagonal-Green's-function}\\
\Gamma^{\dagger}({\bf k},\omega)&=&-{\bar{\Delta}_{\rm h}({\bf k})\over [\omega-\xi_{\bf k}
-\Sigma^{({\rm h})}_{1}({\bf k},\omega)][\omega+\xi_{\bf k}+\Sigma^{({\rm h})}_{1}({\bf k},-\omega)]
-\bar{\Delta}^{2}_{\rm h}({\bf k})},
\end{eqnarray}
\end{subequations}
\end{widetext}
respectively, where $\xi_{\bf k}=Zt\chi_{1}\gamma_{{\bf k}}-Zt'\chi_{2}\gamma_{{\bf k}}'-\mu$ is the mean-field
charge carrier spectrum with the spin correlation functions $\chi_{1}=\langle S^{+}_{i}S^{-}_{i+\hat{\eta}}\rangle$
and $\chi_{2}=\langle S_{i}^{+}S_{i+\hat{\tau}}^{-}\rangle$,
$\gamma_{{\bf k}}=(1/Z)\sum_{\hat{\eta}}e^{i{\bf k}\cdot\hat{\eta}}$,
$\gamma_{{\bf k}}'=(1/Z)\sum_{\hat{\tau}}e^{i{\bf k}\cdot\hat{\tau}}$, and $Z$ is the number of the nearest-neighbor
or next nearest-neighbor sites on a square lattice. The effective charge carrier pair gap
$\bar{\Delta}_{\rm h}({\bf k})=\Sigma^{({\rm h})}_{2}({\bf k},\omega=0)$, therefore it is closely related to the
charge carrier self-energy $\Sigma^{({\rm h})}_{2}({\bf k},\omega)$ in the {\it particle-particle channel}
\cite{feng0306}, and can be expressed explicitly as a d-wave form
$\bar{\Delta}_{\rm h}({\bf k})=\bar{\Delta}_{\rm h}\gamma^{({\rm d})}_{{\bf k}}$ with
$\gamma^{({\rm d})}_{{\bf k}}=({\rm cos} k_{x}-{\rm cos}k_{y})/2$, while the effective charge carrier pair gap
parameter $\bar{\Delta}_{\rm h}$ and the charge carrier self-energy $\Sigma^{({\rm h})}_{1}({\bf k},\omega)$ in the
{\it particle-hole channel} have been given in Ref. \onlinecite{feng12}. In this case, the Fermi energy is determined
by the pole of the Green's function (\ref{hole-Green's-function}). In particular,
$\Sigma^{({\rm }h)}_{1}({\bf k},\omega)$ can be written as \cite{feng12}
$\Sigma^{({\rm }h)}_{1}({\bf k},\omega)=[2\bar{\Delta}_{\rm pg}({\bf k})]^{2}/[\omega+M_{\bf k}]$ with the effective
normal-state pseudogap $\bar{\Delta}_{\rm pg}({\bf k})$ and the energy spectrum $M_{\bf k}$, which reflects that the
effective normal-state pseudogap $\bar{\Delta}_{\rm pg}({\bf k})$ originates from
$\Sigma^{({\rm }h)}_{1}({\bf k},\omega)$ \cite{feng12}. In this case, the imaginary part of
$\Sigma^{({\rm }h)}_{1}({\bf k},\omega)$ can be obtained as
${\rm Im}\Sigma^{({\rm }h)}_{1}({\bf k},\omega)=-\pi [2\bar{\Delta}_{\rm pg}({\bf k})]^{2}\delta(\omega+M_{\bf k})$.
This imaginary part is defined as a momentum dependence of the characteristic scattering rate (then the characteristic
lifetime), and the type of this imaginary part is an intrinsic property of the electronic state of cuprate
superconductors, since it has been used to provide a robust fit for the electron spectrum and spatially inhomogeneous
differential conductances \cite{Vishik09}.

It has been shown that the d-wave charge carrier pairing state in the particle-particle channel originating from the
kinetic energy by exchanging the spin excitations also leads to form a d-wave electron Cooper pairing state
\cite{feng0306}. For discussions of the electronic structure in the SC-state, we need to calculate the electron
diagonal Green's function $G(i-j,t-t') =\langle\langle C_{i\sigma}(t); C^{\dagger}_{j\sigma}(t')\rangle \rangle$, which
is a convolution of the spin Green's function $D^{(0)-1}({\bf k},\omega)=(\omega^{2}-\omega^{2}_{\bf k})/B_{\bf k}$
and charge carrier diagonal Green's function (\ref{hole-diagonal-Green's-function}) in the CSS fermion-spin theory
\cite{guo07}, and can be evaluated in terms of the spectral representation as,
\begin{widetext}
\begin{eqnarray}\label{electron-Green's-function}
G({\bf k},\omega)&=&{1\over 2N}\sum_{{\bf p}}{B_{{\bf p}+{\bf k}}\over \omega_{{\bf p}+{\bf k}}}
\int_{-\infty}^{+\infty}{d\omega'\over 2\pi}A_{\rm h}({\bf p},\omega')\left [{n_{\rm F}(\omega')
+n_{\rm B}(\omega_{{\bf p}+{\bf k}})\over \omega+\omega'-\omega_{{\bf p}+{\bf k}}}
+{1+n_{\rm B}(\omega_{{\bf p}+{\bf k}})-n_{\rm F}(\omega')\over\omega+\omega'+\omega_{{\bf p}+{\bf k}}}
\right ], ~~~~~~
\end{eqnarray}
\end{widetext}
where the charge carrier spectral function $A_{\rm h}({\bf k},\omega)=-2{\rm Im}g({\bf k},\omega)$, while the spin
excitation spectrum $\omega_{\bf k}$ and the function $B_{\bf k}$ have been given in Ref. \onlinecite{guo07}. This
convolution of the spin Green's function and charge carrier Green's function reflects the charge-spin recombination
\cite{anderson91}. The electronic state in a solid is characterized by its energy dispersion as well as the
characteristic lifetime (then the characteristic scattering rate) of an electron placed into such a state. This state
is just represented by the electron Green's function (\ref{electron-Green's-function}), while the electron spectral
function $A({\bf k},\omega)$ is directly related to the analytically continued electron Green's function
(\ref{electron-Green's-function}) as $A({\bf k},\omega)=-2{\rm Im}G({\bf k},\omega)$, and can be obtained explicitly
in terms of the electron Green's function (\ref{electron-Green's-function}) as,
\begin{eqnarray}\label{spectral-function}
A({\bf k},\omega)&=&{1\over 2N}\sum_{{\bf p}}{B_{{\bf p}+{\bf k}}\over \omega_{{\bf p}+{\bf k}}}
\{ A_{\rm h}({\bf p},\omega_{{\bf p}+{\bf k}}-\omega)[n_{\rm F}(\omega_{{\bf p}+{\bf k}}-\omega)\nonumber\\
&+&n_{\rm B}(\omega_{{\bf p}+{\bf k}})]+A_{\rm h}({\bf p},-\omega_{{\bf p}+{\bf k}}-\omega)\nonumber\\
&\times&[1+n_{\rm B}(\omega_{{\bf p}+{\bf k}})-n_{\rm F}(-\omega_{{\bf p}+{\bf k}}-\omega)]\}.
\end{eqnarray}
In particular, this spectral function (\ref{spectral-function}) is measurable via the ARPES technique and can provide
an important information about quasiparticle excitations \cite{Damascelli03}.

\section{Particle-hole asymmetry electronic state due to the presence of the normal-state pseudogap}
\label{electronic-structure}

In this section, we discuss the particle-hole asymmetry electronic state in cuprate superconductors and the related
doping and temperature dependence of the Fermi arc length in the presence of the normal-state pseudogap. In our recent
work \cite{feng12}, we have shown that there is a coexistence of the SC gap and normal-state pseudogap in the whole SC
dome, where the effective charge carrier pair gap parameter $\bar{\Delta}_{\rm h}$ has a domelike shape of the doping
dependence, while the magnitude of the effective normal-state pseudogap parameter $\bar{\Delta}_{\rm pg}$ is
particularly large in the underdoped regime, then it smoothly decreases with increasing doping, and eventually
disappears together with superconductivity at the end of the SC dome. In particular, this $\bar{\Delta}_{\rm pg}$ is
closely related to the charge carrier quasiparticle coherent weight $Z_{\rm hF}$ as \cite{feng12}
$Z^{-1}_{\rm hF}=1+[2\bar{\Delta}_{\rm pg}({\bf k})]^{2}/M^{2}_{\bf k}|_{{\bf k}=[\pi,0]}$, this is why $Z_{\rm hF}$
grows linearly with doping \cite{guo07,feng0306}. We have therefore established a relationship between
$\bar{\Delta}_{\rm pg}$ and $Z_{\rm hF}$, and the obtained evolution of both $\bar{\Delta}_{\rm pg}$ and $Z_{\rm hF}$
with doping is qualitative agreement with the corresponding experimental data \cite{Hufner08,dlfeng00,ding01}. Since
the SC-state in the kinetic energy driven SC mechanism is controlled by both the SC gap and quasiparticle coherence
\cite{feng0306}, and in this sense, the normal-state pseudogap is a necessary ingredient for superconductivity.
Moreover, $\bar{\Delta}_{\rm pg}$ is also temperature dependent, and it vanishes at the normal-state pseudogap
crossover temperature $T^{*}$, while this $T^{*}$ as a function of doping has a similar doping dependence as
$\bar{\Delta}_{\rm pg}$ \cite{feng12}. Furthermore, by calculation of the ratio of the effective charge carrier pair
gap parameter $\bar{\Delta}_{\rm h}$ and the charge carrier pair gap parameter $\Delta_{\rm h}$, we \cite{feng12} have
also obtained the coupling (interaction) strength $V_{\rm eff}$, and the result shows clearly that $V_{\rm eff}$
smoothly decreases upon increasing doping from a strong-coupling case in the underdoped regime to a weak-coupling side
in the overdoped regime. In particular, we \cite{feng12} find that all $T^{*}$, $\bar{\Delta}_{\rm pg}$, and
$V_{\rm eff}$ show the same trend with doping, i.e., $T^{*}\sim \bar{\Delta}_{\rm pg}\sim V_{\rm eff}$, while such a
relationship among $T^{*}$, $\bar{\Delta}_{\rm pg}$, and $V_{\rm eff}$ has been observed experimentally on cuprate
superconductors \cite{Kordyuk10}.

\subsection{Doping and temperature dependence of the Fermi arc length}\label{Fermi-arc}

\begin{figure}[h!]
\includegraphics[scale=0.45]{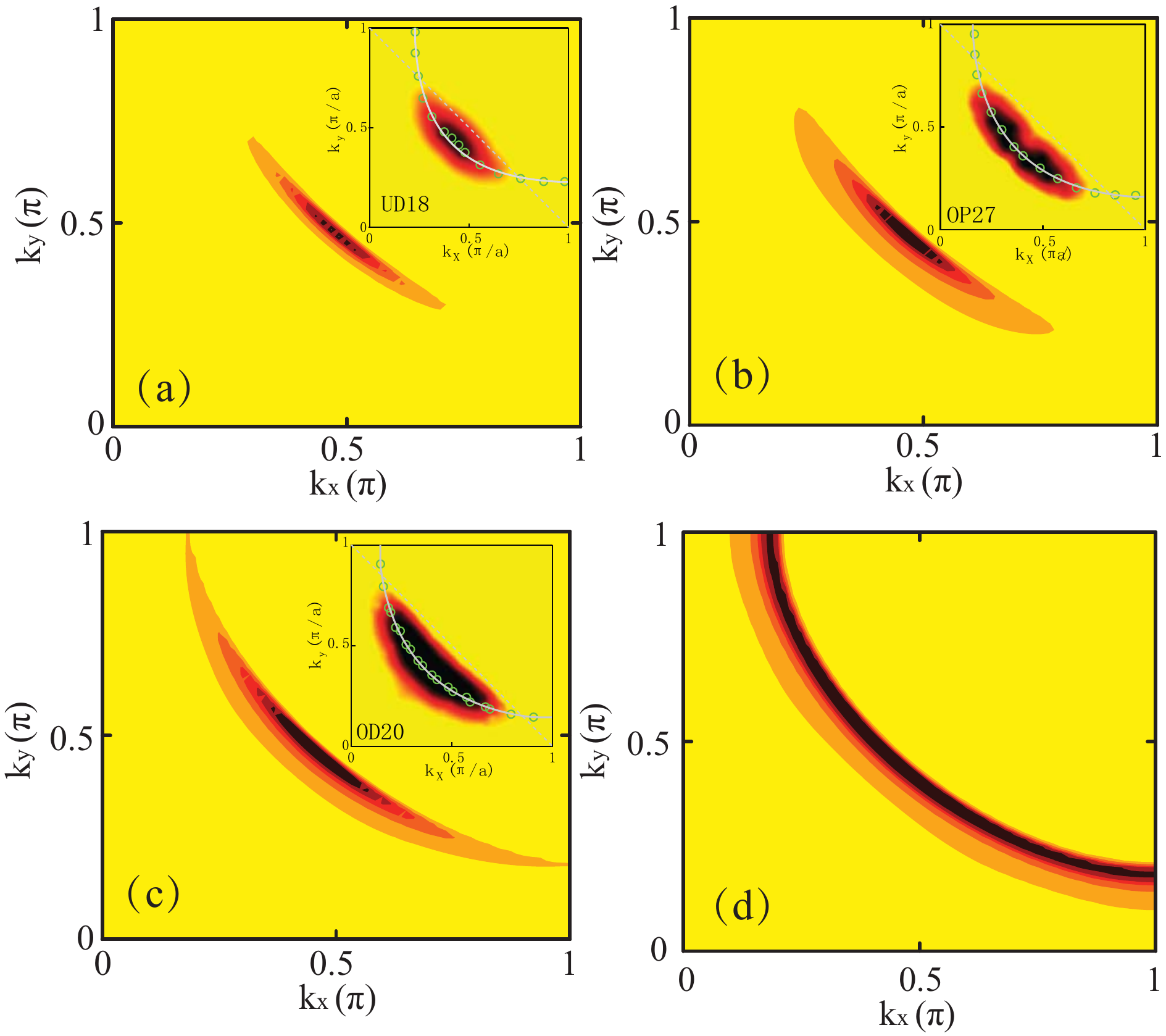}
\caption{(Color online) The doping evolution of the spectral intensity maps at the Fermi energy for (a) $\delta=0.09$,
(b) $\delta=0.15$, (c) $\delta=0.21$, and (d) $\delta=0.26$ with $t/J=2.5$, $t'/t=0.3$, and $J=110$meV in $T=2.55$K.
Inset: the corresponding experimental data of Ca$_{2-x}$Na$_{x}$CuO$_{2}$Cl$_{2}$ at the underdoping, the optimal
doping, and the overdoping, respectively, taken from Ref. \onlinecite{Meng11}. \label{fig1}}
\end{figure}

\begin{figure}[h!]
\includegraphics[scale=0.55]{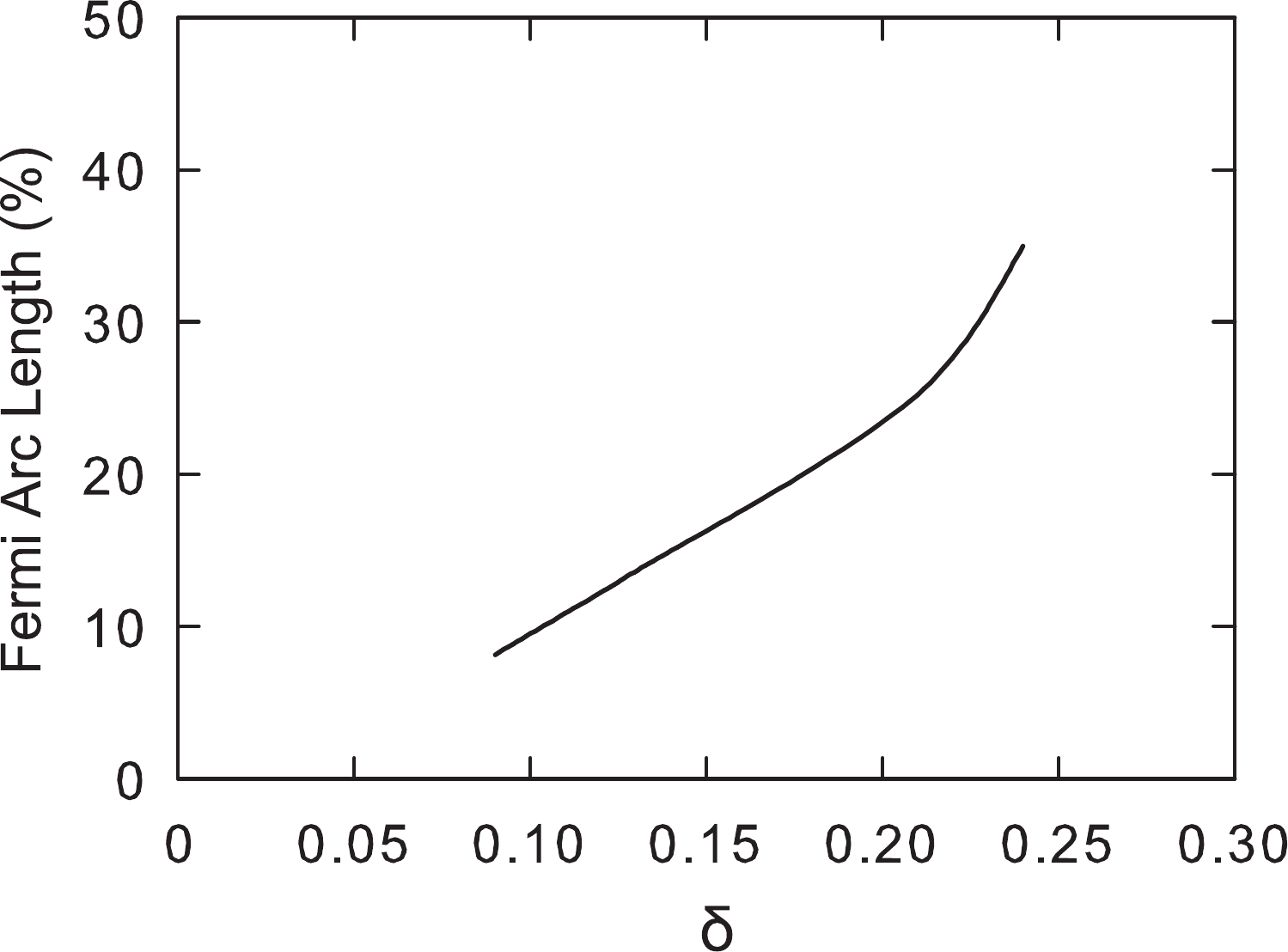}
\caption{The Fermi arc length as a function of doping in $T=2.55$K with $t/J=2.5$, $t'/t=0.3$, and $J=110$meV. On the
$y$ axis, 0\% is the node and 100\% is the antinode. \label{fig2}}
\end{figure}

In the following discussions, we discuss the evolution of the Fermi arc length with doping and temperature. The notion
of the Fermi surface is one of the characteristic concepts in the field of condensed matter physics, and it plays a
crucial role in the understanding of interacting electron systems. In Fig. \ref{fig1}, we show the maps of the spectral
intensity at the Fermi energy for (a) the underdoping $\delta=0.09$, (b) the optimal doping $\delta=0.15$, (c) the
overdoping $\delta=0.21$, and (d) the doping near the end of the SC dome $\delta=0.26$ with parameters $t/J=2.5$,
$t'/t=0.3$, and $J=110$meV in temperature $T=2.55$K. For comparison, the corresponding experimental results
\cite{Meng11} of Ca$_{2-x}$Na$_{x}$CuO$_{2}$Cl$_{2}$ for the underdoping, the optimal doping, and the overdoping are
also shown in Fig. \ref{fig1}(a), Fig. \ref{fig1}(b), and Fig. \ref{fig1}(c), respectively (inset). It is shown clearly
that our present theoretical results capture all qualitative features of the doping dependence of the Fermi arc length
observed experimentally on cuprate superconductors \cite{Meng11}. In the underdoped regime, the Fermi surface does not
form a continuous contour in momentum space, instead, it is broken into disconnected arcs. However, this Fermi arc
increases its length as a function of doping, and there is a tendency towards to form a continuous contour in momentum
space. This tendency is particularly obvious in the overdoped regime, where the Fermi arc becomes flat, and then it
covers the full length of the Fermi surface at the doping near the end of the SC dome [see Fig. \ref{fig1}(d)], in
qualitative agreement with experimental data \cite{Meng11}. Moreover, we have fitted our present results, and found
that the Fermi arc length rise almost linearity with doping. This anticipated result is plotted in Fig. \ref{fig2}.

\begin{figure}[h!]
\includegraphics[scale=0.7]{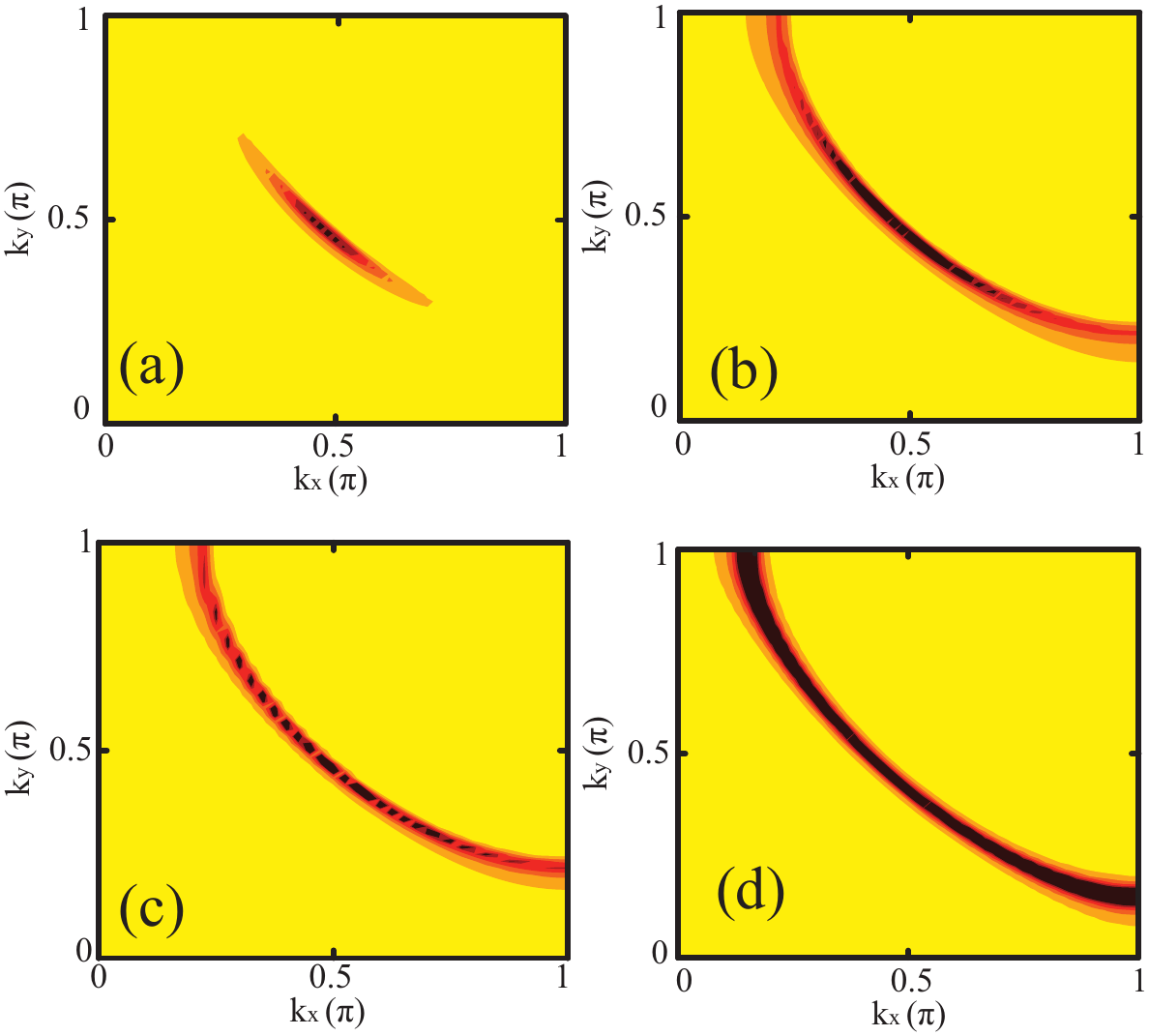}
\caption{(Color online) The temperature evolution of the spectral intensity maps at the Fermi energy for $\delta=0.09$
in (a) $T=2.55$K, (b) $T=57.45$K, (c) $T=70.2$K, and (d) $T=242.55$K with $t/J=2.5$, $t'/t=0.3$, and $J=110$meV.
\label{fig3}}
\end{figure}

\begin{figure}[h!]
\includegraphics[scale=0.52]{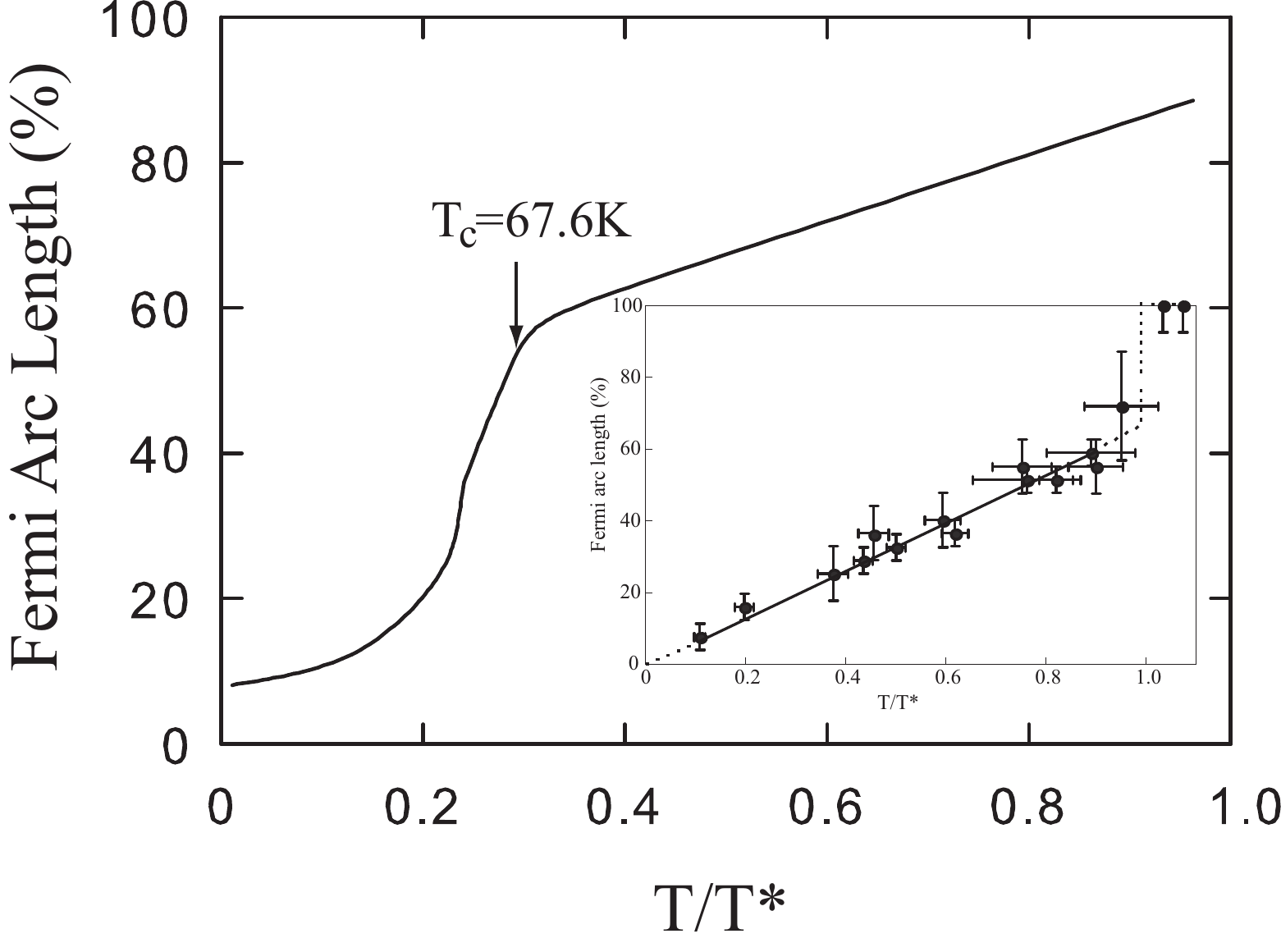}
\caption{The Fermi arc length as a function of temperature for $\delta=0.09$ with $t/J=2.5$, $t'/t=0.3$, and
$J=110$meV. Inset: the corresponding experimental result of Bi$_{2}$Sr$_{2}$CaCu$_{2}$O$_{8+\delta}$ taken from
Ref. \onlinecite{Kanigel06}. On the $y$ axis, 0\% is the node and 100\% is the antinode. \label{fig4}}
\end{figure}

To analyze the evolution of the spectral intensity at the Fermi energy with temperature, we have made a series of
calculations for the spectral density at the Fermi energy with different temperatures, and the results of the maps of
the spectral density at the Fermi energy for $\delta=0.09$ with $t/J=2.5$, $t'/t=0.3$, and $J=110$meV in (a) $T=2.55$K,
(b) $T=57.45$K, (c) $T=70.2$K, and (d) $T=242.55$K are plotted in Fig. \ref{fig3}. Within the kinetic energy driven SC
mechanism, the calculated $T_{\rm c}=67.6$K and $T^{*}=238.7$K, respectively, for $\delta=0.09$. For $T<<T_{\rm c}$,
the Fermi arc shrinks to the nodal region [see Fig. \ref{fig3}(a)]. However, with increasing temperatures, in
particular, for $T>T^{*}$, the Fermi surface forms a continuous contour in momentum space [see Fig. \ref{fig3}(d)]. To
show this temperature dependence of the Fermi arc length clearly, the result for the extracted Fermi arc length as a
function of temperature for $\delta=0.09$ with $t/J=2.5$, $t'/t=0.3$, and $J=110$meV is plotted in Fig. \ref{fig4} in
comparison with the corresponding experimental result \cite{Kanigel06} of Bi$_{2}$Sr$_{2}$CaCu$_{2}$O$_{8+\delta}$
(inset). Our result shows that the Fermi arc length is a linear function of temperature in the temperature range
$T_{\rm c}<T<T^{*}$, i.e., it increases linearly with increasing temperatures, which is in good agreement with
experimental data \cite{Norman98,Kanigel06,Nakayama09} in this temperature range. However, below $T_{\rm c}$, although
the Fermi arc length monotonically increases up to $T_{\rm c}$, the Fermi arc length is a nonlinear function of
temperature. We have also noted that a possible nonlinear temperature dependence of the Fermi arc length seems to have
been observed \cite{Nakayama09} experimentally on (Bi,Pb)$_{2}$Sr$_{2}$CuO$_{6}$.

\subsection{Particle-hole asymmetry electronic state}\label{asymmetry}

\begin{figure}[h!]
\includegraphics[scale=0.62]{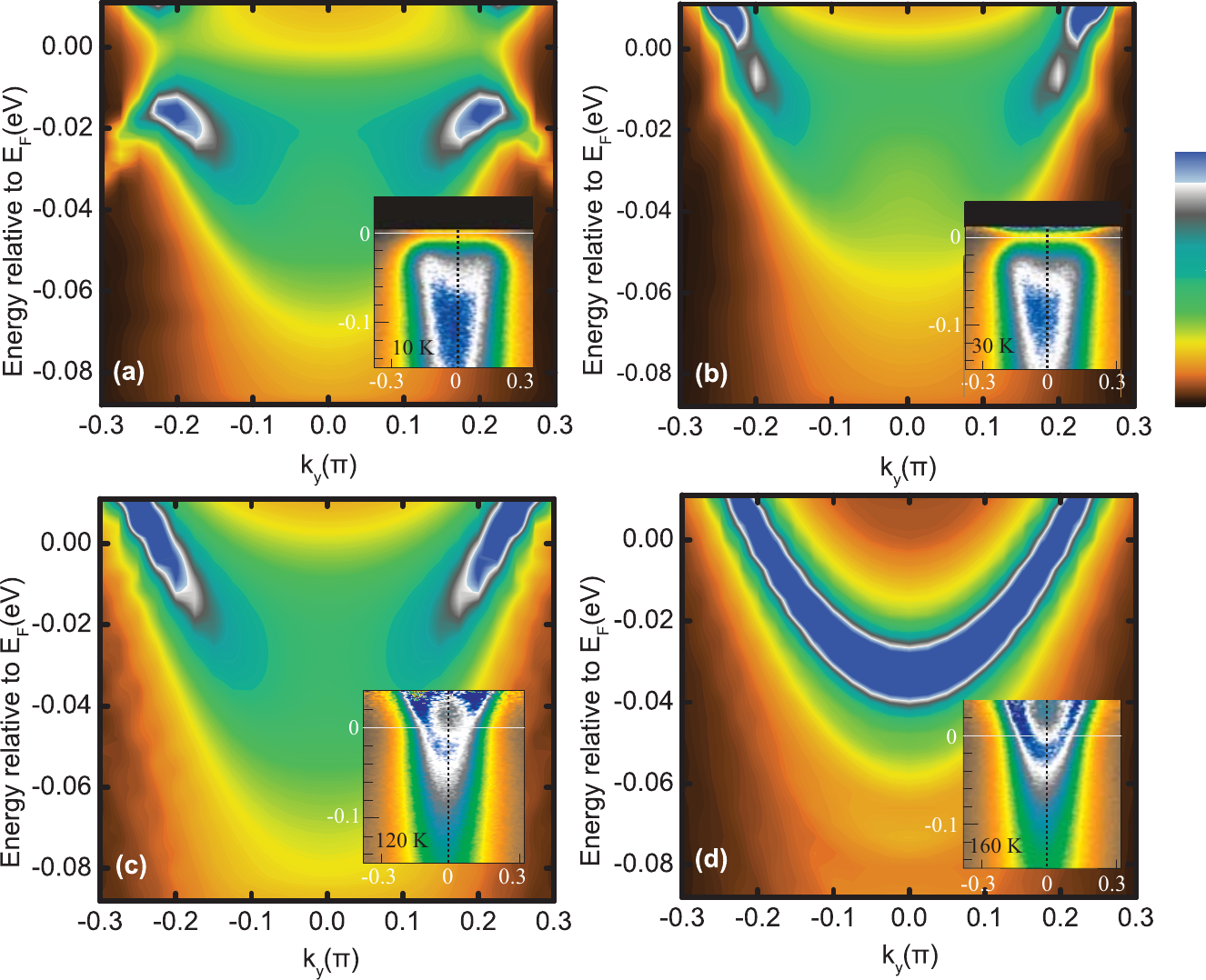}
\caption{(Color online) The maps of the electron spectral function as a function of momentum along the direction
$[\pi,-\pi] \rightarrow [\pi,0]\rightarrow [\pi,\pi]$ for $\delta=0.09$ with $t/J=2.5$, $t'/t=0.3$, and $J=110$meV in
(a) $T=2.55$K, (b) $T=57.45$K, (c) $T=70.2$K, and (d) $T=242.55$K. Inset: the corresponding experimental results of
Pb$_{0.55}$Bi$_{1.5}$Sr$_{1.6}$La$_{0.4}$CuO$_{6+\delta}$ taken from Ref. \onlinecite{Hashimoto10}. \label{fig5}}
\end{figure}

\begin{figure}[h!]
\includegraphics[scale=0.7]{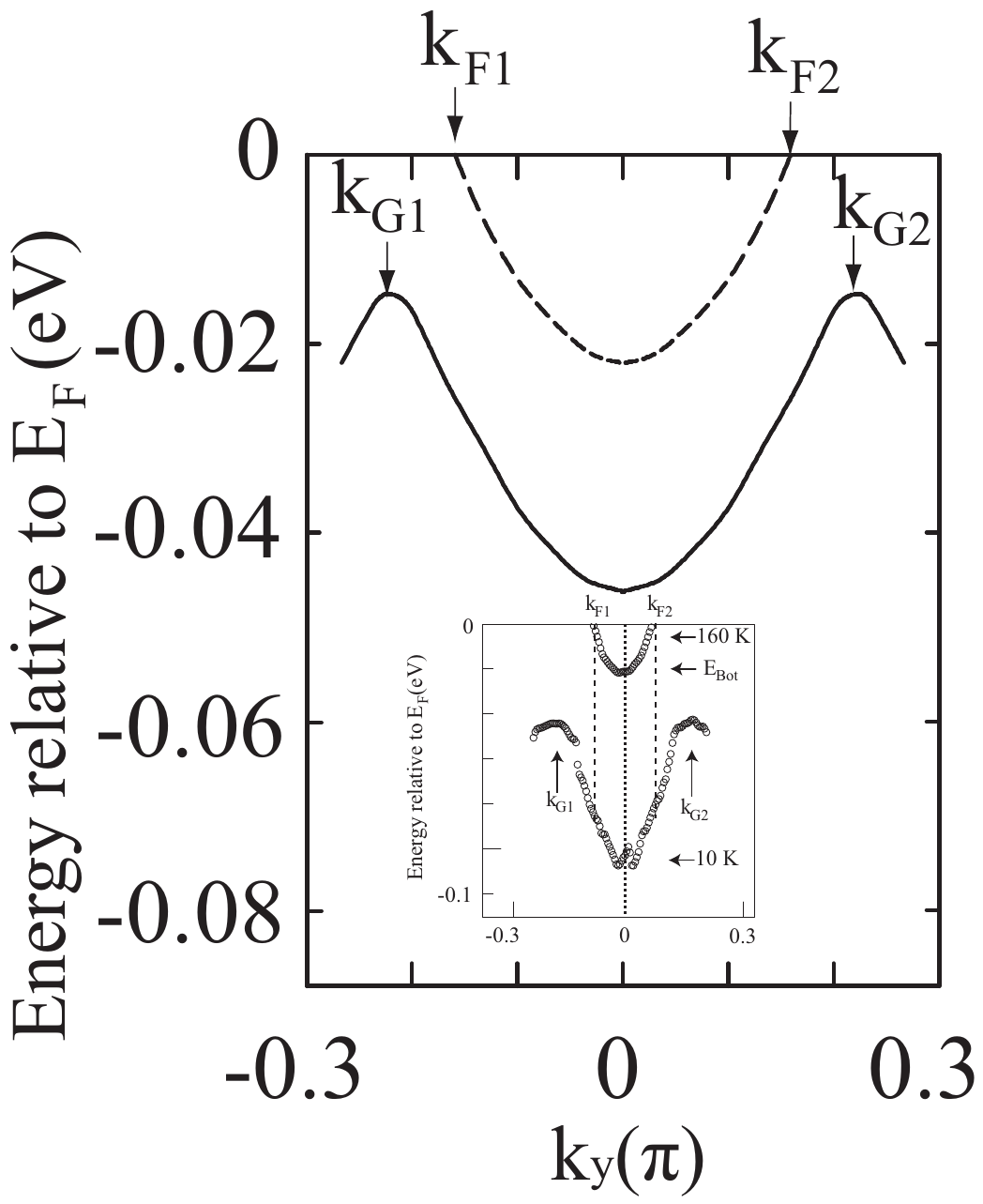}
\caption{The positions of the lowest energy quasiparticle peaks in the electron spectral function as a function of
momentum along the direction $[\pi,-\pi] \rightarrow [\pi,0]\rightarrow [\pi,\pi]$ for $\delta=0.09$ with $t/J=2.5$,
$t'/t=0.3$, and $J=110$meV in $T=2.55$K (solid line) and $T=242.55$K (dashed line). Inset: the corresponding
experimental results of Pb$_{0.55}$Bi$_{1.5}$Sr$_{1.6}$La$_{0.4}$CuO$_{6+\delta}$ taken from Ref.
\onlinecite{Hashimoto10}. \label{fig6}}
\end{figure}

Now we turn to discuss the particle-hole asymmetry electronic state in the presence of the normal-state pseudogap. In
the conventional superconductors \cite{schrieffer83,Hashimoto10}, a particle-hole symmetric energy gap opens from the
normal-state dispersion because of the homogeneous superconductivity, where there is an alignment between the Fermi
crossing $k_{\rm F}$ and the back-bending or the saturation momentum of the dispersion in the gapped states. However,
the recent improvements in the resolution of the ARPES experiments allowed for an experimental verification of the
particle-hole symmetry breaking in the normal-state pseudogap state of cuprate superconductors \cite{Hashimoto10,He11},
in particular, this particle-hole asymmetry electronic state observed from the ARPES experiments is manifested itself
by the nonalignment between the Fermi crossing $k_{\rm F}$ and the back-bending or the saturation momentum of the
dispersion in the gapped states. For a better understanding of the nature of the particle-hole asymmetry electronic
state in cuprate superconductors \cite{Hashimoto10,He11}, we have performed a calculation for the spectral function
$A({\bf k},\omega)$ (\ref{spectral-function}) along the direction $[\pi,-\pi] \rightarrow [\pi,0]\rightarrow [\pi,\pi]$
in different temperatures, and the results of the maps of $A({\bf k},\omega)$ as a function of momentum for
$\delta=0.09$ with $t/J=2.5$, $t'/t=0.3$, and $J=110$meV in (a) $T=2.55$K, (b) $T=57.45$K, (c) $T=70.2$K, and (d)
$T=242.55$K are plotted in Fig. \ref{fig5} in comparison with the corresponding experimental data \cite{Hashimoto10}
of Pb$_{0.55}$Bi$_{1.5}$Sr$_{1.6}$La$_{0.4}$CuO$_{6+\delta}$ (inset). For $T>T^{*}$, the spectrum as a function of
momentum has a parabolic dispersion of the intensity maximum with two clear Fermi level crossings at $k_{\rm F1}$ and
$k_{\rm F2}$ momenta and a bottom reaching $E_{\rm bot}\sim 0.02$eV at $[\pi,0]$ point. In this case, the normal-state
of the system is a Landau Fermi liquid similar to that of an ordinary metal. However, for $T<T^{*}$, the spectral
weight centroid is transferred towards a higher binding energy, where although the spectral weight of the quasiparticle
excitation is strongly suppressed at low temperatures, the intensity maximum of the spectrum can be defined and traced
as function of momentum. In particular, this characteristic feature of the spectrum is still persists even in the
SC-state ($T<T_{\rm c}$). To show this point clearly, we plot the positions of the low energy quasiparticle peaks in
$A({\bf k},\omega)$ as a function of momentum along the direction $[\pi,-\pi]\rightarrow [\pi,0]\rightarrow [\pi,\pi]$
for $\delta=0.09$ with $t/J=2.5$, $t'/t=0.3$, and $J=110$meV in $T=2.55$K (solid line) and $T=242.55$K (dashed line)
in Fig. \ref{fig6} in comparison with the corresponding experimental results \cite{Hashimoto10} of
Pb$_{0.55}$Bi$_{1.5}$Sr$_{1.6}$La$_{0.4}$CuO$_{6+\delta}$ (inset). Obviously, for $T>T^{*}$, the spectrum is gapless
at the Fermi crossings $k_{\rm F1}$ and $k_{\rm F2}$, and the quasiparticle peaks at low energies disperse
parabolically with momentum. However, for $T<T^{*}$, the dispersion at the relatively low energy shows the back-bending
at $k_{\rm G1}$ and $k_{\rm G2}$ momenta, while the band bottom at $[\pi,0]$ point has been pushed far away from the
normal-state $E_{\rm bot}$ appeared for $T>T^{*}$. In this case, the spectrum is fully gapped, and no back-bending
appears at $k_{\rm F1}$ and $k_{\rm F2}$. These results are in qualitative agreement with these observed from ARPES
experimental measurements \cite{He11,Hashimoto10} on Pb$_{0.55}$Bi$_{1.5}$Sr$_{1.6}$La$_{0.4}$CuO$_{6+\delta}$.

\begin{figure}[h!]
\includegraphics[scale=0.55]{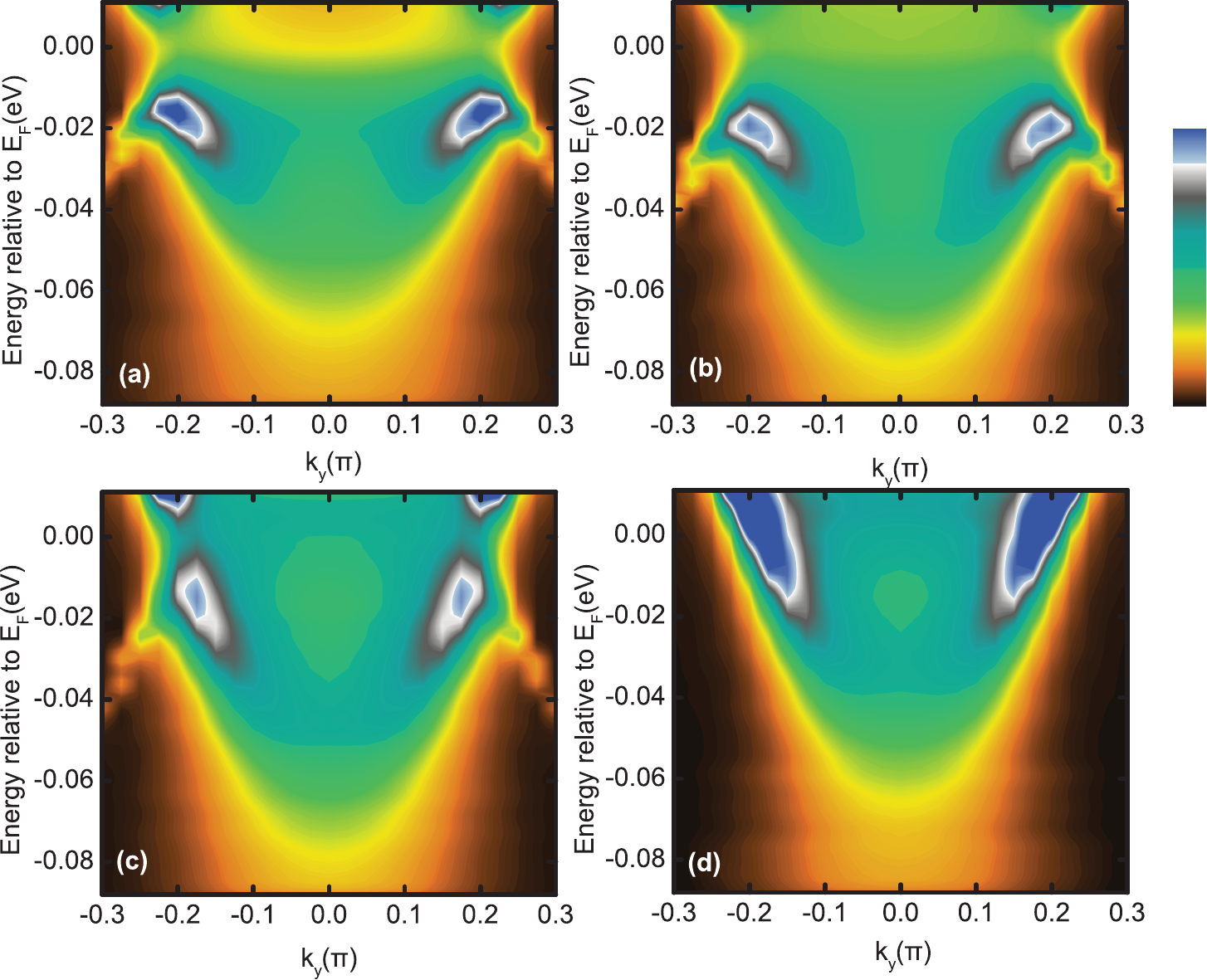}
\caption{(Color online) The maps of the electron spectral function as a function of momentum along the direction
$[\pi,-\pi] \rightarrow [\pi,0]\rightarrow [\pi,\pi]$ for (a) $\delta=0.09$, (b) $\delta=0.15$, (c) $\delta=0.21$, and
(d) $\delta=0.26$ with $t/J=2.5$, $t'/t=0.3$, and $J=110$meV in $T=2.55$K.\label{fig7}}
\end{figure}

\begin{figure}[h!]
\includegraphics[scale=0.65]{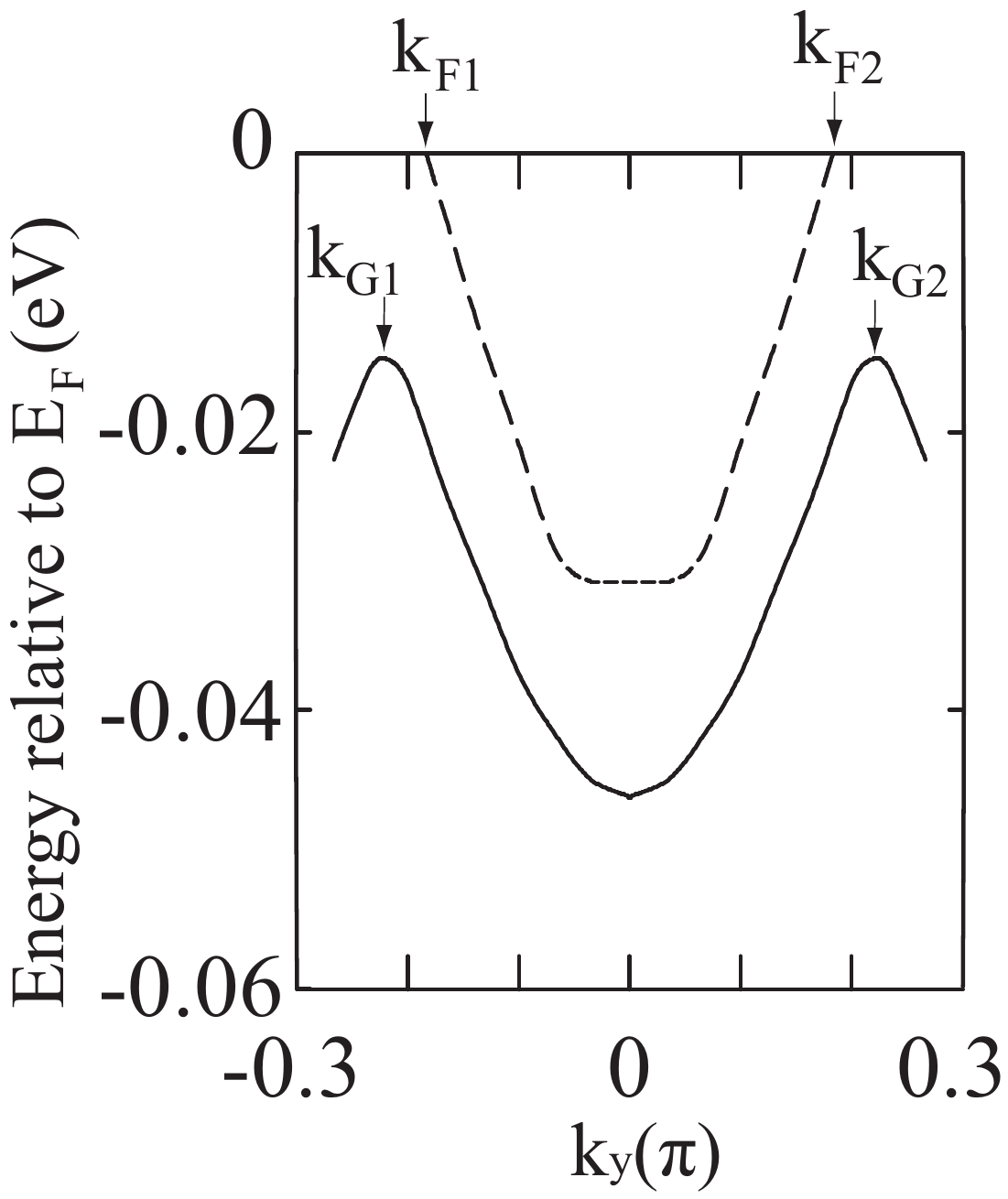}
\caption{The positions of the lowest energy quasiparticle peaks in the electron spectral function as a function of
momentum along the direction $[\pi,-\pi] \rightarrow [\pi,0]\rightarrow [\pi,\pi]$ for $\delta=0.09$ (solid line) and
$\delta=0.26$ (dashed line) with $t/J=2.5$, $t'/t=0.3$, and $J=110$meV in $T=2.55$K. \label{fig8}}
\end{figure}

Furthermore, we have also discussed the doping dependence of the particle-hole asymmetry electronic state. In Fig.
\ref{fig7}, we plot the maps of the spectral function $A({\bf k},\omega)$ as a function of momentum along the direction
$[\pi,-\pi] \rightarrow [\pi,0]\rightarrow [\pi,\pi]$ for (a) $\delta=0.09$, (b) $\delta=0.15$, (c) $\delta=0.21$, and
(d) $\delta=0.26$ with $t/J=2.5$, $t'/t=0.3$, and $J=110$meV in $T=2.55$K, where the behavior of the evolution of the
particle-hole asymmetry electronic state with doping is very similar to that of the doping dependence of the
particle-hole asymmetry electronic state shown in Fig. \ref{fig5}. For the doping near the end of the SC dome, the
maxima of the spectral function have an approximately parabolic dispersion [see Fig. \ref{fig7}(d)]. In this case, the
quasiparticle excitations are gapless, and the dispersion crosses the Fermi level at two momenta $k_{\rm F1}$ and
$k_{\rm F2}$. However, the dispersion rises to a minimum binding energy and then bends back in the underdoping [see
Fig. \ref{fig7}(a)], where there is an energy gap in the quasiparticle excitations. In particular, this back-bending
occurs at two momenta $k_{\rm G1}$ and $k_{\rm G2}$, which are separated from the corresponding Fermi crossing points
$k_{\rm F1}$ and $k_{\rm F2}$, as summarized in Fig. \ref{fig8}. Moreover, the gap size and the distance between the
$k_{\rm G1}$ and $k_{\rm F1}$ (then $k_{\rm G2}$ and $k_{\rm F2}$) decrease with increasing doping.

A natural question is why the the particle-hole asymmetry electronic state in cuprate superconductors and the related
doping and temperature dependence of the Fermi arc length can be described qualitatively within the framework of the
kinetic energy driven superconductivity \cite{feng0306}. The reason is that the emergence of the doping and temperature
dependence of the normal-state pseudogap \cite{feng12}. This follows a fact that in the framework of the kinetic energy
driven SC mechanism, besides the pairing mechanism in the particle-particle channel provided by the charge carrier
interaction directly from the kinetic energy by exchanging spin excitations, this same microscopic interaction also
induces a normal-state pseudogap state in the particle-hole channel \cite{feng12} as mentioned above. This normal-state
pseudogap opens at the Fermi energy for $T<T^{*}$, which competes with superconductivity by suppressing the spectral
weight of the quasiparticle excitation. Since the magnitude of the normal-state pseudogap decreases with increasing
doping and temperature \cite{feng12}, this leads to that the strength of the suppression for the spectral weight also
decreases with increasing doping and temperature. In this case, the particle-hole asymmetry electronic state and the
related doping and temperature dependence of the Fermi arc length are a natural consequence due to the presence the
normal-state pseudogap in the particle-hole channel. In the extremely low temperatures ($T\sim 0$), as a result of the
strong suppression for the spectral weight in the underdoped regime, the electronic state breaks the particle-hole
symmetry, where the majority contribution for the electron spectrum comes from the nodal region, i.e., the most
low-energy states are located around the nodal region, forming the Fermi arcs. However, the strength of this
suppression decreases with increasing doping, which leads to that the Fermi arc length increases with increasing
doping. In particular, for the doping near the end of the SC dome, the normal-state pseudogap is negligible
\cite{feng12}, i.e., $\bar{\Delta}_{\rm pg}\approx 0$. In this case, the full charge carrier diagonal and off-diagonal
Green's functions (\ref{hole-Green's-function}) are reduced as a simple d-wave BCS formalism \cite{zhao12},
\begin{subequations}\label{BCSform}
\begin{eqnarray}
g({\bf k},\omega)&=&{U^{2}_{{\rm h}{\bf k}}\over\omega-E_{{\rm h}{\bf k}}}+{V^{2}_{{\rm h}{\bf k}}\over\omega+
E_{{\rm h}{\bf k}}}, \\
\Gamma^{\dagger}({\bf k},\omega)&=&-{\bar{\Delta}_{\rm h}({\bf k})\over 2E_{{\rm h}{\bf k}}}\left ( {1\over\omega-
E_{{\rm h}{\bf k}}}-{1\over\omega + E_{{\rm h}{\bf k}}}\right ),
\end{eqnarray}
\end{subequations}
although the pairing mechanism in the particle-particle channel is driven by the kinetic energy by exchanging spin
excitations, where the charge carrier quasiparticle coherence factors
$U^{2}_{{\rm h}{\bf k}}=(1+\xi_{{\bf k}}/E_{{\rm h}{\bf k}})/2$ and
$V^{2}_{{\rm h}{\bf k}}=(1-\xi_{{\bf k}}/E_{{\rm h}{\bf k}})/2$, and the charge carrier quasiparticle spectrum
$E_{{\rm h}{\bf k}}=\sqrt{\xi^{2}_{{\bf k}}+\mid\bar{\Delta}_{\rm h}({\bf k})\mid^{2}}$. It is thus similar to the
conventional superconductors \cite{schrieffer83,Hashimoto10}, in this simple d-wave BCS formalism (\ref{BCSform}),
the Fermi arc covers the full length of the Fermi surface, and an alignment between the Fermi crossing $k_{\rm F}$
and the back-bending in the electronic state appears. This is why the SC coherence of the low energy quasiparticle
excitation and the related physical properties in the heavily overdoped regime are well described by a simple d-wave
BCS formalism \cite{Lee07,guo07,zhao12}. On the other hand, in the normal-state pseudogap phase ($T_{c}<T<T^{*}$),
the charge carrier pair gap $\bar{\Delta}_{\rm h}=0$, and then the full charge carrier diagonal Green's function
(\ref{hole-Green's-function}) can be reduced as \cite{feng12,zhao12},
\begin{eqnarray}\label{unusual-metal-form}
g({\bf k},\omega)={\alpha^{(n)}_{1{\bf k}}\over\omega-E^{+}_{{\rm h}{\bf k}}}
+{\alpha^{(n)}_{2{\bf k}}\over\omega-E^{-}_{{\rm h}{\bf k}}},
\end{eqnarray}
where the charge carrier quasiparticle coherence factors $\alpha^{(n)}_{1{\bf k}}=(E^{+}_{{\rm h}{\bf k}}+M_{\bf k})
/(E^{+}_{{\rm h}{\bf k}}-E^{-}_{{\rm h}{\bf k}})$ and $\alpha^{(n)}_{2{\bf k}}=-(E^{-}_{{\rm h}{\bf k}}+M_{\bf k})
/(E^{+}_{{\rm h}{\bf k}}-E^{-}_{{\rm h}{\bf k}})$, while the dispersion has two branches, $E^{+}_{{\rm h}{\bf k}}=
[\xi_{{\bf k}}-M_{\bf k}+\sqrt{(\xi_{{\bf k}}+M_{\bf k})^{2}+16\bar{\Delta}^{2}_{\rm pg}({\bf k})}]/2$ and
$E^{-}_{{\rm h}{\bf k}}=[\xi_{{\bf k}}-M_{\bf k}-\sqrt{(\xi_{{\bf k}}+M_{\bf k})^{2}
+16\bar{\Delta}^{2}_{\rm pg}({\bf k})}]/2$. In this normal-state pseudogap phase, the particle-hole symmetry is broken
due to the presence the normal-state pseudogap. However, with increasing temperatures, in particular, for $T>T^{*}$,
both the normal-state pseudogap $\bar{\Delta}_{\rm pg}=0$ and the charge carrier pair gap $\bar{\Delta}_{\rm h}=0$,
and in this case, the full charge carrier diagonal Green's function (\ref{hole-Green's-function}) is reduced as,
\begin{eqnarray}\label{metal-form}
g({\bf k},\omega)&=&{1\over\omega-\xi_{{\bf k}}}.
\end{eqnarray}
In this case, the normal-state is a particle-hole symmetry Landau Fermi liquid similar to that of an ordinary metal,
and the Fermi arc turns into a continuous contour in momentum space.

\section{Conclusions}\label{conclusions}

In this paper, we have studied the particle-hole asymmetry electronic state in cuprate superconductors and the related
doping and temperature dependence of the Fermi arc length within the framework of the kinetic energy driven SC
mechanism. By taking into account the interplay between the SC gap and normal-state pseudogap, we have reproduced
qualitatively the essential feature of the evolution of the Fermi arc length with doping and temperature. Our results
show that the particle-hole asymmetry electronic state in cuprate superconductors is a natural consequence due to the
presence the normal-state pseudogap in the particle-hole channel. The Fermi arc length increases with increasing
temperature below $T^{*}$, and it covers the full length of the Fermi surface for $T>T^{*}$. In particular, in analogy
to the temperature dependence of the Fermi arc length, the low-temperature Fermi arc length in the underdoped regime
increases with increasing doping, and then it evolves into a continuous contour in momentum space near the end of the
SC dome. The theory also predicts an almost linear doping dependence of the Fermi arc length, which should be verified
by further experiments. Since the knowledge of the particle-hole asymmetry electronic state is of considerable importance
as a test for the microscopic SC mechanism in cuprate superconductors, the qualitative agreement between the present
theoretical results and ARPES experimental data also provides an important confirmation of the nature of the SC phase
of cuprate superconductors as a coexistence of the kinetic energy driven d-wave SC-state in the particle-particle
channel and the normal-state pseudogap state in the particle-hole channel in the whole SC dome.

\acknowledgments

This work was supported by the funds from the Ministry of Science and Technology of China under Grant Nos. 2011CB921700
and 2012CB821403, and the National Natural Science Foundation of China under Grant No. 11074023.

\end{document}